\documentclass[aps,twocolumn,,superscriptaddress,showpacs,amssymb, prl]{revtex4}

\usepackage{amsbsy,latexsym}
\usepackage{amsfonts}
\usepackage{amssymb}
\usepackage[mathscr]{eucal}

\newcommand{\be}{\begin{eqnarray}}
\newcommand{\ee}{\end{eqnarray}}
\newcommand{\one}{\mbox{$1 \hspace{-1.0mm}  {\bf l}$}}

\newcommand{\ket}[1]{\left|{#1}\right\rangle}
\newcommand{\bra}[1]{\left\langle{#1}\right|}

\newcommand{\ketbrad}[1]{\left|{#1}\rangle\!\langle{#1}\right|}
\newcommand{\ketbra}[2]{\left|{#1}\rangle\!\langle{#2}\right|}
\newcommand{\mean}[1]{\langle{#1}\rangle}
\def\bea{\begin{eqnarray}}
\def\eea{\end{eqnarray}}

\newcommand{\ea}{\emph{et al.}}

\newcommand{\sumint}[1] {\sum_{\begin{array}{c}\\[-1.2em] \scriptstyle#1\end{array}}\kern-1.3em\int}
\def\C{\hbox{$\mit I$\kern-.7em$\mit C$}}
\def\N{\hbox{$\mit I$\kern-.3em$\mit N$}}

\def\tr{{\rm tr}}

\usepackage{dcolumn} 
\usepackage{bm} 

\usepackage{epsfig}

\usepackage{amsmath,amsfonts,amssymb,graphics,graphicx,epsfig,color,times,bbm}

\begin{document}

\title{How to hide a secret direction}

\author{E. Bagan}
\affiliation{Grup de F\'isica Te\`orica \& IFAE, Universitat Aut\`onoma de Barcelona, Spain} 
\author{J. Calsamiglia}
\affiliation{Grup de F\'isica Te\`orica \& IFAE, Universitat Aut\`onoma de Barcelona, Spain} 
\author{ R. Demkowicz-Dobrza\'nski}
\affiliation{Center for Theoretical Physics, Polish Academy of Sciences, Poland} 
\author{R. Munoz-Tapia}
\affiliation{Grup de F\'isica Te\`orica \& IFAE, Universitat Aut\`onoma de Barcelona, Spain}


\begin{abstract}
We present a procedure to share a secret spatial direction in the absence of a common reference frame using a multipartite quantum state. The procedure guarantees that the parties can determine the direction if they perform joint measurements on the state, but fail to do so if they restrict themselves to local operations and classical communication (LOCC). We calculate the fidelity for joint measurements, give bounds on the fidelity achievable by LOCC, and prove that there is a non-vanishing gap between the two of them, even in the limit of infinitely many  copies. The robustness of the procedure under particle loss is also studied.  As a by-product  we find bounds on the probability of discriminating by LOCC between the invariant subspaces of total angular momentum $N/2$ and $N/2-1$ in a system of $N$ elementary spins.
\end{abstract}


\maketitle

A good number of quantum information protocols that take advantage of the laws of quantum mechanics to keep secrets in different scenarios have been put forward over the last years.  Quantum key distribution~\cite{gisin02}, which is probably the most prominent example, allows two parties to establish a secret random key. Using quantum secret sharing protocols~\cite{qsectret} one can share secret information (classical or quantum) among several parties, so that  the so-called authorized set can perfectly unveil it whereas any other set of parties cannot to any extent.  Finally, quantum data hiding~\cite{terhal01} allows also for  the possibility to share information (classical or quantum) among many parties, but with the promise that they can only unveil it by performing joint operations, i.e. any local strategy assisted with classical communication will reveal (almost) nothing.  

Here we present a procedure that uses spin-$1/2$ particles to share a secret direction  in a similar fashion: the parties can unveil it only if they perform joint measurements on all particles. Three observations are in order here. First, the estimation of the direction will always be limited by the quantum nature of measurements, and thus a perfect estimation of the secret direction is only possible in the limit of an infinite number of particles. Second, as in quantum data hiding, and in contrast to quantum secret sharing, the information that can be obtained by LOCC, although negligible,  is not strictly zero.  
Last, the information shared in quantum data hiding or secret sharing is abstract in that it can be represented in terms of qubits or bits disregarding their particular physical support. In contrast, a direction contains information of a very particular sort: it is \emph{physical information}. A direction is an intrinsic property of some physical systems.  Similarly, codifying a particular direction on an $N$-body state $\rho$ in the absence of a reference frame  involves a very specific operation, namely  a rigid rotation $U^{\otimes N} \rho\, U^{\dagger \otimes N}$ (see e.g. work on establishing common reference frames~\cite{frames}). Hence, protocols for direction sharing are distinct to those of quantum data hiding for which there are no constraints on how the information is encoded.

Most quantum communication protocols require that the parties involved have a common reference frame (see, however, \cite{bartlett03}). A quantum direction sharing protocol could be a primitive to establish such common frames without compromising the security requirements of the communication protocol.

A central ingredient, and a shared motivation, to this work is quantum state estimation. The topics of quantum state estimation and discrimination~\cite{estimation,gill00}
are arguably among the first quantum information theoretical problems
with an important impact on other  research areas in quantum information ranging from entanglement theory~\cite{horodecki03}, to state disturbance and teleportation~\cite{banaszek01} or quantum cloning~\cite{bruss98}.  Also, very recently Bacon \ea~\cite{bacon05} have found new efficient quantum algorithms by recasting some computational problems in terms of state estimation.  

The usual scenario in state estimation is to consider~$N$ copies of a given unknown state and study the performance of different protocols (according to some figure of merit)  as a function of~$N$. The action of general unitaries on qubits is the same as that of the physical rotations on spin~$1/2$ systems. Hence, all the results regarding the fidelity of a state estimation protocol under collective, local fixed and local adaptive measurements can be applied in the context of direction estimation. A common conclusion from these results is that all protocols achieve perfect determination in the limit of infinite number of copies, i.e. the fidelity (see below for a precise definition) is $F=1-{\cal O}(N^{-1})$.
In this paper we build on recent work~\cite{demko05},
where the effect of correlations on the fidelity of 
permutation-invariant states is calculated and its relation to universal cloning machines is dicussed.
We provide exact results for joint measurements and non-trivial bounds for LOCC measurements.
We find the surprising result that for local protocols the fidelity does not reach its optimal value ($F=1$) even when an infinite number of copies is available. However, when joint measurements are used, the very same state provides the fidelity of a perfect ``gyroscope'', i.e. $N$ spins pointing in the same direction.

In order to illustrate the direction hiding protocol we imagine the following fictitious scenario:
A space station sends its (large) corps of space explorers to look for resources in other galaxies. The official in charge of the operation wants to make sure that the whole corps sticks together. To this end, he provides every explorer with a single spin~$1/2$ particle, and with a set of instructions that specify how to obtain the direction home, which has been encoded in the quantum state of the spins. 
Upholding the principle of equality, he also choses the state to be permutation-invariant. 
The essential property of the state is that only by performing joint measurements on it  one can retrieve the direction home with accuracy, thus forcing the corps to stick together. If a significant fraction of it is left behind or refuses to join the rest, then the corps will not be able to make it back home. At the same time if a small fraction is captured by unfriendly forces, the rest of the corps will still be able to decode the direction faithfully, while the enemy will not be able to learn the whereabouts of the space station.

{\bf Direction hiding state:} Here we give an $N$-spin state that encodes the unknown direction, denoted with the unit Bloch vector~$\vec n$, in a way that it can be estimated perfectly by joint measurements, but extremely poorly by LOCC measurements.  We use the fidelity $F$ as a figure of merit to quantify the explorers' ability to determine the direction,
$F=\mean{f}\equiv(1+\Delta)/2$ with $f=(1+\vec{n}\cdot\vec{n}_{\chi})/2 \equiv (1+\Delta_{n,\chi})/2 $,
where~$\vec{n}_{\chi}$  is the guess that they make for~$\vec n$ after obtaining the measurement outcome $\chi$. The average is taken over all possible states and over all measurement outcomes.  
  
An arbitrary permutationally invariant state can be written~as (we assume $N$ even for simplicity)
$$
\rho=\!\sum_{j=0}^{J}\sum_{\alpha=1}^{n_j}\rho_{(j,\alpha)};\; \rho_{(j,\alpha)}\!=\kern-.3cm\sum_{m,m'=-j}^{j}\kern-.3cm\lambda^{(j)}_{m,m'}\ketbra{j,m,\alpha}{j,m',\alpha} ,
$$
where $J=N/2$, $j$ is the total angular momentum, $m$ is its projection along the encoded direction~$\vec{n}$, and $n_{j}$ is the multiplicity of the spin-$j$ representation. 
Given a generalized measurement, represented by a positive operator-valued measure (POVM) ${\cal M}=\{M_{\chi}\}$, 
the expected fidelity can be written as~\cite{bagan06},
\begin{equation}
\Delta=\sum_{j}n_{j}\left(\sum_{m}\frac{m}{j+1} \lambda^{(j)}_{mm}\right)\left(\sum_{m} \frac{m}{j} \frac{\Omega^{(j)}_{mm}}{2j+1}\right) ,
\label{eq:Delta}
\end{equation}
 where $\Omega^{(j)}_{mm'}=\bra{j,m,\alpha}\Omega\ket{j,m,\alpha}$, $\Omega=\sum_{\chi} U(\vec n_{\chi})^\dagger M_{\chi} U(\vec n_{\chi})$, and $U(\vec n_{\chi})$ is the unitary associated with the rotation that brings the quantization axis $\vec{z}$ to the estimated direction $\vec{n_{\chi}}$.

The completeness of the POVM implies $\sum_{m}\Omega^{(j)}_{mm}/(2j+1)\kern-.1em=\kern-.1em1$. From here it follows that the optimal fidelity is given by
$\Delta=\sum_{j}n_j|\sum_{m}m \lambda^{(j)}_{m,m}|/(j+1)$.
For a fixed single-particle purity $r$ [i.e., $\rho_1=(\openone+r\,\vec n\cdot\vec\sigma)/2$, 
where~$\vec\sigma$ is a vector made out of the three Pauli matrices], this expression is maximized by $\lambda^{(J)}_{m,m} =p \,\delta_{Jm}$  and $\lambda^{(J\!-\!1)}_{m,m} =(1-p)\, \delta_{J\!-\!1,-m}/(N-1)$. This corresponds to the state:
\begin{equation}
\rho=\!p \ketbrad{\vec{n}}^{\otimes N}\!\!+\!\frac{1-p}{N_{\sigma}}\!\sum_{\sigma} \ketbrad{-\vec{n}}^{\otimes N-2}\kern-.2em\otimes \ketbrad{\psi^-} ,
\label{eq:Prep}
\end{equation}
where the sum is taken over all permutations of the position of the singlet $\ket{\psi^-}$, and where the value of~$p$ determines the purity of the local state.
In particular, one can take~$p= (N-2)/(2N-2)$ so that every individual spin is maximally mixed ($r=0$) and contains no information about the direction. In spite of that, perfect direction estimation is still possible (asymptotically): 
\be
\Delta= p\frac{N}{N+2}+ (1\!-\! p)\frac{N-2}{N}\approx 1-\frac{2}{N} .
\label{eq:Deltacol}
\ee
The  direction hiding state \eqref{eq:Prep} encodes $\vec n$ as efficiently as $N$ parallel spins. In the following we will prove that one can only reach this maximum fidelity under joint measurements, i.e., for any LOCC strategy the fidelity does not reach one even in the strict  limit~$N\rightarrow \infty$. This means that, no matter how large~$N$ is, if the space explorers do not perform collective measurements, with high probability they will {\em not} be able to return home. We note in passing that the proposed state and measurement scheme can be implemented efficiently~\cite{bacon}.

{\bf LOCC upperbound:} To obtain an upper bound to the~LOCC fidelity we further assume that the space explorers are told about the specific preparation~(\ref{eq:Prep}) used in the communication protocol:  they know they are given a state with total angular momentum~$j=J$ pointing along~$\vec n$ or with $j=J-1$ pointing along~$-\vec n$, with probabilities~$p$ or $1-p$ respectively. We can view $J$ and $J-1$ as tags that  tell the explorers to move forward or backwards relative to the guessed directions, which hopefully will be close to $\vec n$ or $-\vec n$, respectively.
The explorers' ability to estimate the secret direction is conditioned to their~ability to read  these tags, or more precisely, to their ability to discriminate between spin-$J$ and spin-($J-1$) states. 
Hence,  they have to optimize their measurements to give both a good estimate of the direction in which the state is pointing, and also a low error probability of discrimination between the two possible tags (``forward" if $j=J$ vs. ``backwards" if $j=J-1$). We will obtain a bound for the fidelity focusing on this second aspect, i.e. discrimination, while neglecting the difficulties of estimating the direction.

We divide the set ${\cal M}$ of POVM elements in two groups, ${\cal F}$ and ${\cal B}$ (after ``forward" and ``backwards"), those for which the final guess and the measured direction coincide and those for which they are opposite  (corresponding to the case were the explorers judge that the input state had total angular momentum $J-1$). It is clear that the wrong assignments will result in a negative contribution to $\Delta$ ---at least in average. We can thus obtain a rough upper bound by assuming that (i) $\Delta_{n,\chi}=1$ when the explorers correctly identify the signal state, and (ii) $\Delta_{n,\chi}=0$ if they fail to do so. 
In the asymptotic limit, $N\!\rightarrow\!\infty$, we have that $p=1/2$, and $\Delta$ is thus bounded by the probability $p_{S}$ of discriminating between the~$J$ and $J-1$ subspaces:
\begin{eqnarray}
\Delta&\leq& 
 \frac{1}{2}\left[ \sum_{M_\chi\in{\cal F}} \tr\left(M_\chi\frac{\one_{J}}{d_{J}}\right) +\!\!\sum_{M_\chi\in{\cal B}} \tr\left(M_\chi\frac{\one_{J\!-\!1}}{d_{J\!-\!1}}\right)\right]\nonumber\\
&=&\frac{1}{2}\left[\frac{1}{d_{J}}\tr( Q_{J} \one_{J})+\frac{1}{d_{J\!-\!1}}\tr( Q_{J\!-\!1} \one_{J\!-\!1})\right]=: p_{S},
\label{eq:ub}
\end{eqnarray}
where $\openone_j$ stands for the identity restricted to the spin-$j$ subspace, whose dimension is~$d_{j}=n_{j}\times (2j+1)$.  In obtaining~(\ref{eq:ub}), we have used Schur's lemma and implicitly defined the local discrimination POVM  ${\cal Q}=\{Q_{J},Q_{J-1}\}$. 
This bound is rough, but of course is saturated for collective POVM since in this case one can perfectly discriminate between the spin-$J$ and spin-$(J-1)$ subspaces. Notice also that the bound will not be tight for local measurements  since typically one cannot avoid having negative contributions to the fidelity when the explorers fail to discriminate between those subspaces.

Let us now give a bound on $p_{S}$, i.e. on the probability that $Q_{J}$ and $Q_{J\!-\!1}$ project successfully on the spin-$J$ and spin-$(J-1)$ subspaces respectively.
We first notice that any optimal POVM can be locally symmetrized by the action of the rotation 
group. The weights of every POVM element on each invariant subspace will remain untouched under this action. Hence, for our purposes we can now take
$$
Q_{J}= a \one_{J}+b \one_{J\!-\!1}+c \one_{j<J\!-\!1};\quad Q_{J\!-\!1}=\openone-Q_J.
$$

The action of~$\cal Q$, if it ought to be LOCC, cannot produce entanglement. By performing~$\cal Q$ on a particular separable state and imposing that the resulting state cannot be entangled we
find certain constraints on~$a$ and~$b$, which in turn will give a bound on the discrimination probability.
The state we choose is a four-party  pure state~$\ket{\psi}^{1234}$ which is separable with respect to the partition~$\{(13),(24)\}$, but maximally entangled with respect to the partition 
$\{(12),(34)\}$. More precisely,
$\ket{\psi}^{1234}\!=\!\ket{\psi^{+}}^{13}\ket{\psi^{+}}^{24}$,
where $\ket{\psi^{\pm}}^{\mu\nu}=(\ket{{0}}^{\mu}\ket{{1}}^{\nu}\!\pm\!\ket{{1}}^{\mu}\ket{{0}}^{\nu})/\sqrt{2}$, with $\ket{{k}}\! =\!\ket{J/2,J/2-k}$.
By performing the LOCC measurement~$\cal Q$ on parties~$12$, which we denote by~${\cal Q}^{12}$, one cannot create entanglement between parties~$34$. The state of the latter conditioned to having obtained the outcome~$J$ has the form $\tr_{12}[Q_{J}^{12}\ket{\psi}^{1234}\!\bra{\psi}]=a \rho^{34}_J+b \rho^{34}_{J\!-\!1}+c \rho^{34}_{J\!-\!2}$, where $\tr_{12}$ is the partial trace over parties~$12$ ---and analogous expression is obtained for the outcome~$J-1$.
Using the partial transposition criteria on these states we obtain necessary conditions for them to be separable. 
Maximizing~$p_{S}=(1+a-b)/2$ subject to these conditions we find an upper bound for the fidelity $\Delta< p_{S}\leq 2-\sqrt{5}/2=0.88<1$ which holds
for all $N$.
Larger values for~$\Delta$ would necessarily imply generation of entanglement by the LOCC protocol we have just discussed. 

\medskip{\bf LOCC lowerbounds:} 
The point of showing the previous bound is that it proves the existence of a \emph{finite} gap even in the limit \mbox{$N\rightarrow\infty$}. We will now present some results that suggest that the gap is in fact very large, corresponding to a LOCC fidelity that is lower than that of a single spin-1/2 pure state. We will do this by studying a sound  family of LOCC estimation protocols. Strictly speaking this will only give a lower bound to the LOCC fidelity. However, as it will become apparent below, this is arguably the maximum fidelity attainable by LOCC  in the asymptotic limit.

It is by now clear that one needs to learn whether the direction is enconded in the $J$ or $J\!-\!1$ subspace, i.e. a good protocol has to be able to detect the presence of a singlet in an otherwise fully symmteric state. It is also clear that if the axis~$\pm \vec n$ is known, one can detect the presence of the singlet by measuring each spin along~$\pm \vec n$  (if all but one of  the outcomes are identical the signal state is in the $J-1$ subspace). The measurement $\{M(\vec m, x)\}$ we propose consist of a two step process where we first try to determine the axis~$\pm\vec n$ of the encoded direction by performing measurements on the first~$N_{0}$ spins. In a second step, one measures the projections of the remaining $N_{1}=N-N_{0}$  spins along the estimated axis ($\pm\vec m$) so as to detect the presence of the singlet.
Using this scheme, or any other LOCC scheme, one cannot reach $F=1$ mainly for two reasons: (i) 
the singlet state may involve one or two of the $N_0$ spins of the first step and thus it may pass unnoticed;
(ii) the axis can only be estimated within a precision of $(\Delta \theta)^2=4/N_{0}$, which blurs the effect of a singlet at the second measurement step.

We formalize the above strategy by a POVM on $[{\mathbb C}^2]^{\otimes N_0}\otimes [{\mathbb C}^2]^{\otimes N_1}$ that strictly speaking is semi-local. This will enable us to obtain a closed form for the fidelity
without actually giving up locality in the asymptotic limit, since this POVM 
can be realized  by LOCC with arbitrary accuracy as $N_{0}\!\rightarrow\!\infty$ (see below). The POVM elements are given by, 
$
M(\vec m, x)=O(\vec m)\otimes E(\vec m,x)
$
where $O(\vec m)=(N_{0}+1)[J_{0},J_{0}]_{\vec m}+(N_{0}-1)\sum_{\alpha=1}^{N_{0}-1}[J_{0}-1,J_{0}-1,\alpha]_{\vec m}$ defines an optimal (and covariant) POVM to estimate the axis, but does not reveal the total angular momentum.  We have introduced the notation $J_0=N_0/2$,  $[\phi]_{\vec m}=U(\vec{m}) \ketbrad{\phi} U(\vec{m})^\dagger $ where $\ket{\phi}=\ket{j,m,\alpha}$. The operators  $E(\vec m,x)$, acting on the last $N_{1}$ particles, correspond to measuring  the projection of the spin along $\vec m$ on each of the particles, where $x$ is the total number of  ``up" outcomes. It can also be written as:
$E(\vec{m},x)=[J_1,J_1-x]+\sum_{\beta=1}^{N_1-1}[J_1-1,J_1-x,\beta]+\dots$.
The guess associated to $M(\vec m, x)$ is $(-1)^{f(x)}\vec m$, i.e. $\vec m$ determines the axis of the explorers'  guess, while the function $f$, from $\{0,1,\dots,N_1\}$ to $\{0,1\}$, determines its orientation or, in other words, their guess for the tag $j\in \{J,J\!-\!1\}$. 

Using some angular momentum algebra one can calculate the matrix elements of $\Omega$ that are relevant to~(\ref{eq:Delta}). Notice that~$\Omega$ has the simple expression $\Omega =O(\vec{z})\otimes E(\vec z,x)$. With this,
\begin{eqnarray*}
&&\kern-1.em\Delta={N_1!\over2 N!}\sum_x(-1)^{f(x)}{(N-x)!\over(N_1-x)!}
\left\{
{N_0+1\over N+1}
{N-2x\over N+2}\right.\\
&&\kern-1.em-{1\over N(N-1)(N-x)}
       \left[
       (N-2x)\,x {N_0(N_0+1)\over N_1}
       \right. \\
       &&\kern-1.em       \left.\left.
       +
       {N\!\!-\!2x\!-\!2 \over N\!\!-\!x\!-\!1}N(N_0\!-\!\!1)^2\!\! 
 +\!(N\!\!-\!2x)(N_1\!-\!x){N(N_0\!+\!1)\over N_1}
       \right]
\right\},
\end{eqnarray*}
where for simplicity we have taken $p=1/2$. The sum runs from $x=0$ up to $x=N_1$ for all the terms but the very last one, for which $1\le x\le N_1$. In the asymptotic limit this expression is maximized by taking $f(x)\!=\!0$ for $x\!=\!0,N_{1}$, and $f(x)\!=\!1$ otherwise, and by taking $N_1=N_0=N/2$. One obtains
$
\Delta_{\rm max}=1/4+1/(2N)+\mathcal{O}(N^{-2})
$.
Interestingly, we notice that the fidelity decreases with the number  $N$ of spins, which reflects the fact that  it becomes increasingly difficult to detect a singlet among a growing number of parallel spins. 

One can show that if we are given a state $\rho'=q[J,J]_{\vec n}+(1-q)[J\!-\!1,J\!-\!1]_{\vec n}$ and we find a protocol (a POVM $\cal M$) such that
 $\Delta_{J}=1-{ J^{-1}}+\mathcal{O}(J^{-1-a})$ for all~$q$ , where $1\geq a>0$,
then it follows that  $\Omega^{(J)}_{m m}/(2J+1)=\mathcal{O}(J^{-a})$ for $m\neq J$ and $\Omega^{(J)}_{J J}/(2J+1)=1-\mathcal{O}(J^{-a})$. In the limit $N\to\infty$, these matrix elements are exactly those for $\Omega=O(\vec z)$, corresponding to the POVM $O(\vec m)$ above. 
It is easy to verify that the LOCC
state estimation protocol of~\cite{gill00}, which is optimal for pure state estimation, also provides the above asymptotic values for $\Delta_{J}$ when applied to the state~$\rho'$. Hence, we conclude that there exists a LOCC measurement strategy that achieves 
\begin{equation}
\Delta_{ \rm LOCC}={1\over4}+\mathcal{O}(N^{-a}) .
\label{eq:locc}
\end{equation}

This very same protocol can be used to discriminate between the $J$ and $J\!-\!1$ subspaces with probability of success $p_{S}=5/8$. The leading order in~(\ref{eq:locc}) can be easily accounted for by the relation $\Delta_{\rm  LOCC}\approx p_{S}\Delta_{J}-(1-p_{S})\Delta_{J-1}\approx2 p_{S}-1=1/4$, where $\Delta_{J}\approx\Delta_{J-1}\approx1$. This means that an error in discriminating between the invariant subspaces (the tags), has  a very important negative impact on the fidelity. For this reason, the upper bound~(\ref{eq:ub}) is far from being tight.
\begin{figure}[ht]
$$
\begin{picture}(210,115)
\put(5,-10){\includegraphics[width=210pt]{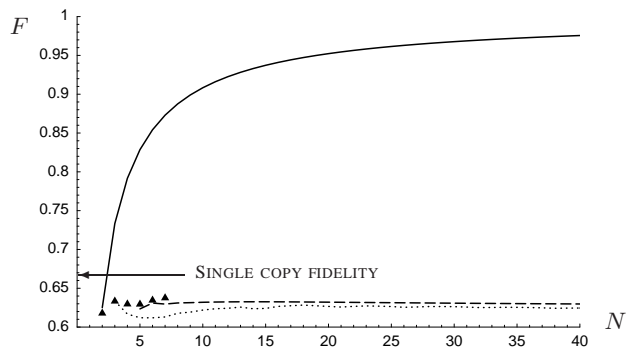}}
\put(-5,110){$F$}
\put(220,-2){$N$}
\linethickness{.2pt} 
\put(61,18.65){\vector(-1,0){40}}
\put(63,17.4){ \sc \scriptsize Single copy fidelity}
\end{picture}
$$
\caption[]{\label{fig:fid} The fidelity for different protocols (see main text): (a) joint measurements (solid); (b) local adaptive projections (triangles); (c) semi-local covariant measurement (dashed); (d) tomography (dotted). 
}
\end{figure}

In Fig.~\ref{fig:fid} we show the fidelity $F=(1+\Delta)/2$ for (a) joint measurements, together with the optimal values for various general protocols: (b) most general sequence of local adaptive projective measurements~\cite{bagan02}, which due to the increasing numerical complexity, is limited to $N\leq 7$; (c) semi-local protocol: perform covariant measurement on~$N/2$ particles to estimate de axis and measure the projection of each of the remaining~$N/2$ spins along it; (d) same as (c), but using standard local tomography to estimate the axis.  We note that all (semi-)local protocols give similar fidelities which, moreover,  fall way below that of a single spin, $F_{1}=2/3$. Note also that tomography performs slightly worse than the semi-local (but asymptotically {\em fully} local) protocol. This can be traced back to the value of the sub-leading term in the asymptotic expression of $\Delta_J$: from~\cite{bagan02} we have $1-\Delta_{J}=\!6/(5 J)$, which is larger than the value $1/J$ required for~(\ref{eq:locc}) to hold.
%

\medskip{\bf Robustness:} 
We finish by studying the robustness of the optimal state when a number~$M$  of parties is left out. A long but straightforward calculation shows that after tracing out a fraction $\xi$ of parties, $\xi=\lim_{N\rightarrow \infty}M/N$, the optimal fidelity can be written as $F=\!1\!-\!\xi/2-\!(1\!-\xi)/\!(N\!-\!M)$. So,  in the case that the fraction of lost parties is vanishingly small ($\xi\!\to\!0$), the fidelity remains the same up to ${\cal O}(N^{-1})$, while it falls below unity already at the leading order for a non-vanishing fraction.

We are grateful to A. Acin, S.~Barnett, S.~Popescu and W.~Wooters, and specially to A.~Monr\`as  for discussions.
We acknowledge financial support from the Spanish MCyT, under Ram\'on y Cajal program (JC) and FIS2005-01369, and CIRIT SGR-00185. RDD acknowledges support from Foundation for Polish Science, the MNiI under PBZ-Min-008/P03/03  and the EC Program QUPRODIS.

\end{document}